\documentclass[fleqn,10pt]{wlscirep}
\title{One-out-of-two Quantum Oblivious Transfer based on Nonorthogonal States}

\usepackage{physics}
\usepackage{multirow}
\usepackage{subfigure}
\usepackage{graphicx}
\usepackage{amsmath}
\usepackage{hyperref}
\usepackage{bookmark}
\usepackage{blindtext}
\usepackage{scrextend}
\usepackage{color}
\newcommand{\tabincell}[2]{\begin{tabular}{@{}#1@{}}#2\end{tabular}}

\author[1,*]{Yao-Hsin Chou}
\author[1]{Guo-Jyun Zeng}
\author[1]{Yu-Shan Yang}
\author[1]{Zhe-Hua Chang}

\affil[1]{Department of Computer Science and Information Engineering, National Chi Nan University, Puli, 54561, Taiwan}
\affil[*]{yhchou@ncnu.edu.tw}
\begin{abstract}
This study proposes a simple and efficient one-out-of-two quantum oblivious transfer (QOT) protocol based on nonorthogonal states. The nonorthogonal property grants quantum bit immunity to some operations in order to achieve the irreversible goal of discarding a message, resulting in a one-out-of-two selection effect. In addition, it can also prevent entangled cheating from an illegal agent. The resulting QOT protocol is therefore built directly on quantum resources, rather than on a two-level structure which must first create two classical keys using the quantum resources (all-or-nothing QOT), and then build the one-out-of-two protocol from there. Furthermore, the proposed protocol allows Alice to test Bob's loyalty by comparing the measurement results. Moreover, the relationship with the no-go theorem is discussed in detail; this relationship is often overlooked in other studies. The no-go theorem discussion and security analysis demonstrate that the proposed protocol does not belong to the no-go theorem constraint, and is secure against both external and internal attacks. In addition, the efficacy analysis shows that the proposed protocol is more efficient than other two-level structure protocols.
\end{abstract}
\begin{document}

\flushbottom
\maketitle
\thispagestyle{empty}

%\noindent Please note: Abbreviations should be introduced at the first mention in the main text – no abbreviations lists. Suggested structure of main text (not enforced) is provided below.

\section*{Introduction}\label{sec:introduction}
%qkd->quantum computer(computational assumptions)->IBM 50bits, google, NASA, and D-WAVE 512 bits->quantum satellite(develop)->more and more feasible->quantum cryptography->ot definition->ot application->ot history(equivalent)->ot development->relationship with Lo's no-go
Since people first began to share information, they have striven toward the ultimate goal of unconditionally secure communication. To this end, cryptographers have used many difficult and complex mathematical problems, for which is very difficult to evaluate all possible solutions. In 1984, Bennett and Brassard\cite{Bennett1984} first proposed quantum key distribution (QKD), also called BB84, based on quantum physics with uncertainty, and the No-Cloning theorem\cite{Wootters1982}, which guarantees unconditionally secure key distribution between a source and a destination\cite{Lo1999,Shor2000,Mayers2001,Koashi2003}. Moreover, the superposition and entanglement properties inspired Deutsch and Jozsa\cite{Deutsch1992}, Shor\cite{Shor1994} and Grover\cite{Grover1996} to develop their parallel search algorithms. These algorithms produced unsafe RSA\cite{Shor1994} and AES\cite{Grover1996,Bernstein2010,Grassl2016} systems, which are famous examples of asymmetric and symmetric cryptography. In addition, quantum computing hardware development is making significant advances. Google, NASA and D-WAVE\cite{Denchev2016} have cooperated to develop an analogous quantum simulator with 512 bits, which is 10 million to 1000 million times more powerful than a single processor core, but cannot launch real quantum algorithms. Meanwhile, IBM has developed a 5-qubit quantum computer, and claims to have broken through the physical limit and hope to be able to develop a 50-qubit quantum computer for commercial use within the next decade. Furthermore, a Chinese research group successfully launched quantum satellites on the 16th of August, 2016\cite{Chin2016}. These satellites create entangled qubits, and are thus able to generate secure keys for globe-spanning quantum communications. In the near future, quantum computers will have a significant effect on our daily lives, as their computational capacity poses a major threat to classical cryptography, while at the same time offering massive potential for far more secure communication.

Since the hardware breakthrough, quantum cryptography has extended and diversified from classical cryptography, has received a great deal more research attention. One important branch is quantum oblivious transfer (QOT), which offers many exciting new applications such as secure computation, bit-commitment, remote coin-flipping, digital contract signing and more. The two most commonly used oblivious transfers (OT) are the all-or-nothing and one-out-of-two protocols. All-or-nothing OT was first introduced by Rabin\cite{Rabin1981} in 1981. In the OT protocol, a sender Alice wants to send a secret message, $m\in\{0, 1\}$, to a receiver Bob who has only a 50\% probability of receiving the message $m$. He could learn either the message $m$ with 100\% reliability, or nothing about $m$. At the end of the OT, Alice remains oblivious as to whether Bob received the message $m$ or not. Subsequently, Even et al.\cite{Even1985} presented one-out-of-two oblivious transfer, in which Alice wishes to transfer two messages, $m_{0}$ and $m_{1}$, to Bob, and he can choose one of them, but will have no idea what the other message is. Analogously, Alice learns nothing about which message Bob chooes when the protocol is complete. In 1988, Cr\'epeau\cite{Crepeau1988} provided a method for building one-out-of-two OT by using p-all-or-nothing OT, in which the receiver has $p$ probability of receiving the message $m$. The receiver can build two key sets, $key_{0}$ and $key_{1}$, which he learns with 100\% certainty and 0\% certainty, as his choices. Based on Bob's choice $j$, he asks Alice to encrypt her messages $m_{0}$ and $m_{1}$ using $key_{0}$ and $key{1}$ if $j=0\Rightarrow key_{0}$, or $j=1\Rightarrow key_{1}$. Then Bob can receive $m_{j}$ by the method. Using this method, Cr\'epeau\cite{Crepeau1988} proved that the two OTs are equivalent at the classical level. Another related founding block is bit-commitment (BC), which was proved computationally equivlent\cite{Bennett1992c,Crepeau2001} with one-out-of-two QOT at the quantum level, which means each can be used as a founding block for the others. These reductions significantly affected OT at both the classical and quantum levels. The discussion in this study will therefore include these improtant precursors. 

Cr\'epeau and Kilian\cite{Crepeau1988a} proposed the first all-or-nothing QOT in 1988, and Bennett et al.\cite{Bennett1992c} proposed the first one-out-of-two QOT protected by quantum error correct code in 1992. In 1994, Cr\'epeau\cite{Crepeau1994} presented a one-out-of-two QOT based on quantum bit-commitment (QBC), which guarantees the security on the assumption that Bob cannot delay the quantum measurement. In 1995, Yao\cite{Yao1995} further proved that the protocol is secure against coherent measurement if QBC is secure. In fact, the possibility of QOT was an open problem\cite{Brassard1996} in 1996. In a period time therefore, the above reductions were used to build other protocols. However, Mayers\cite{Mayers1997} demonstrated that unconditionally secure bit-commitment is impossible. At the same time, Lo and Chau\cite{Lo1997a} took the same result and produced the well known no-go theorem (also called the Mayers-Lo-Chau theorem, the MLC no-go theorem was shown within the scope of non-relativistic physics) in 1997. This means any QBC based QOT is insecure, and implies insecure QOT. Furthermore, Lo\cite{Lo1997} in the same year showed that all one-side two-party computations (allowing only one of the two parties to learn the result) are necessarily insecure, which includes one-out-of-two QOT. This was called Lo's no-go theorem. Therefore, following the above computational equivalent\cite{Crepeau1988,Bennett1992c,Crepeau2001}, these theorems have caused extreme difficulty for the development of QOT research. Finally, however, Colbeck\cite{Colbeck2007} and Buhrman et al.\cite{Buhrman2012} complemented to that two-side two-party computation is insecure in a relativistic quantum protocol.

Recent studies have, however, proposed various methods of avoiding these no-go theorems. In 2002, Shimizu and Imoto\cite{Shimizu2002} presented an interesting communication method analogous to one-out-of-two QOT with a 50\% probability of completing the communication. This means that the receiver cannot learn a message unambigously, and Lo's no-go theorem is thus avoided. They\cite{Shimizu2003} then improved the security of their protocol against entangled pair attacks in 2003. In 2005, Damg$\text{\aa}$rd et al.\cite{Damgard2005} proposed all-or-nothing and bit-commitment protocols based on BB84, which focused on implementation. They noted that no-go theorems did not take three scenarios into account: First, where an agent's computing power is bounded; second, where the communication is noisy; and third, where an eavesdropper is under some physical limitation, e.g., quantum memory is limited. In the same year, Wolf and Wullschleger\cite{Wolf2005} showed a reduction between OT and PR-boxes (proposed by Popescu and Rohrlich\cite{Popescu1997}, also known as a PR non-local box). In 2006, He and Wang\cite{He2006a}, proposed an all-or-nothing QOT using four entangled states, which, as a result, no longer fell into the Lo's no-go theorem\cite{Lo1997} class. Thereafter, they\cite{He2006} showed that two kinds of QOT are nonequivalent at the quantum level, and found that the one-out-of-two QOT built on all-or-nothing QOT using Cr\'epeau's reduction\cite{Crepeau1988} is not rigorously subject to Lo's no-go theorem\cite{Lo1997}. The key point is that the receiver inputs their choice before the sender inputs their message $m_{0}$ and $m_{1}$, and the result is that the function of the one-out-of-two QOT differs from that of Lo's no-go theorem\cite{Lo1997}. Subsequently, Yang et al.\cite{Yang2007} presented a one-out-of-two QOT using tripartite entangled states based Cr\'epeau's reduction\cite{Crepeau1988}, and also showed that it is not covered by the cheating strategy of Lo's no-go theorem\cite{Lo1997}. Unfortunately, the protocol cannot work because it does not change the measurement basis, which means that while Alice can determine of the all-or-nothing QOT key, an eavesdropper will also be able to obtain the key.

So far, researchers have, after the nonequivalence proof\cite{He2006}, developed QOT using different methods and reduction\cite{Crepeau1988}. The methods preposed to date can be roughly classified into 5 types: 1. PR non-locality box; 2. QBC-based QOT; 3. Bit-string QOT; 4. Cr\'epeau's reduction\cite{Crepeau1988} and 5. Others. These methods will be discussed below as follows:
	\begin{enumerate}
		\item \textbf{QBC-based QOT:} The key discussion point here is the MLC no-go theorem\cite{Mayers1997,Lo1997a}, which shows that no QBC is secure, and how it can be used for QOT. In 1999, Kent\cite{Kent1999} proposed an unconditionally secure  BC based on relativistic scenario, which is secure against classical and quantum attacks. He\cite{Kent2003} went on to present a secure quantum bit-string commitment scheme based on redundant code in 2003, which allows the receiver to recover a bit from the bit-string. He also developed unconditionally secure QBC\cite{Kent2012} based on his previous works\cite{Kent1999,Kent2003}. At this point, an interesting question arises, as discussed by He\cite{He2015a} in 2015: could unconditionally secure QBC lead to secure QOT? Although most QBC-based QOT is insecure, He cannot assert whether other kinds of unconditionally secure QOT exist (the transmission He used is Yao's research\cite{Yao1995}), especially relativistic QOT. In addition, He\cite{He2015a} also questions whether Lo and others' no-go theorems\cite{Lo1997,Colbeck2007,Buhrman2012} (includes two-side two-party) covered all kinds of QOT, with proof, such as a class in which the sender helps the receiver to compute a prescribed function. This area clearly requires more research attention and development.%Although, the QBC-based QOT may exist, but most recent QBC-based QOTs doesn't discuss the relationship between the no-go theorem\cite{Mayers1997,Lo1997a}.
		\item \textbf{PR non-locality box:} After the reduction from OT to PR non-locality box by Wolf and Wullschleger\cite{Wolf2005}, Buhrman et al.\cite{Buhrman2006} proposed a QBC protocol based on PR-box, and went on to construct one-out-of-two QOT in 2006. Chou et al.\cite{Chou2010} constructed a PR non-locality box using ten entangled qubits, and used it to build one-out-of-two QOT. Furthermore, Gisinet al.\cite{Gisin2013} introduced a method to build one-out-of-two QOT with an 85.3\% success rate using a single qubit and rotation, much higher than that achieved by Shimizu and Imoto\cite{Shimizu2002}. Despite the PR non-locality property being applied, these protocols did not exhibit the relationship between the no-go theorems; in particular, the security of the two-side two-party method was in doubt\cite{Colbeck2007,Buhrman2012,He2015a}.
		\item \textbf{Bit-string QOT:} Despite the advent of quantum bit-string commitment was developed\cite{Kent2003}, research into bit-string QOT continued. In 2015, Souto et al.\cite{Souto2015} proposed all-or-nothing bit-string QOT based on quantum state computational distinguishability with fully flipped permutations ($\mathrm{QSCD_{\mathit{ff}}}$\cite{Kawachi2012}), which is a hard problem even for a quantum computer. They proved its security against few-qubit coherent attacks, leaving the question of general multi-qubit measurements open. However, He\cite{He2015} noted that Souto's protocol\cite{Souto2015} is not secure, because Bob checks Alice's behavior only when his received message is correct, so a dishonest Alice can always deceive Bob into receiving nothing while ensuring that he remains unaware of the deception. Souto's protocol\cite{Souto2015} also happened to be a building block for constructing one-out-of-two QOT using Cr\'epeau's reduction\cite{Crepeau1988}. Souto et al.\cite{Souto2015a} replied that He's comment\cite{He2015} was not within the scope of the all-or-nothing OT proposed by Rabin\cite{Rabin1981}. Moreover, Souto et al.\cite{Souto2015a} responded to the deception issue with a secure semi-honest one-out-of-two QOT against a malicious Alice. More recently, Plesch et al.\cite{Plesch2017} agreed with He\cite{He2015} that all-or-nothing QOT with security flaws cannot be used to construct a secure one-out-of-two QOT, and presented an improved version which demonstrated that building a secure one-out-of-two QOT requires a perfect all-or-nothing QOT using Cr\'epeau's reduction\cite{Crepeau1988}.
		\item \textbf{Cr\'epeau's reduction\cite{Crepeau1988}:} This category consists mostly of one-out-of-two QOT protocols after He's proof\cite{He2006}, and some one-out-of-two QOT protocols\cite{Yang2013,Yang2014,Yang2015b,Yang2015a,Yang2015c,Yang2017}. The protocol proposed in this study is similar, but does not first generate two keys by all-or-nothing QOT. Li Yang\cite{Yang2013} first presented all-or-nothing QOT by nonorthogonal states, and used it as a basis to construct one-out-of-two QOT in 2013. Meanwhile, Yu-Guang Yang and his research team, in research that began in 2007, have proposed several QOT protocols: They used the same state to detect all cheats, and then build all-or-nothing up to one-out-of-two QOT\cite{Yang2015a}; this lead to an untrusted third party method\cite{Yang2014}, created all-or-nothing QOT using Bell state measurements, and then further built up one-out-of-two QOT\cite{Yang2015c}. Furthermore, they proposed one-out-of-n QOT\cite{Yang2017} based any two nonorthogonal states by cooperative measuring the qubit sequence to create the key, and then build one-out-of-n QOT, in 2017. \textbf{However, these methods have the disadvantage that they require many qubits for the classical key generation before the one-out-of-two QOT. The key point of He's nonequivalence proof\cite{He2006} is that Bob and Alice's inputs should not be independent (The example of He's proof is that Bob choose before Alice inputs her two messages), rather than first creating the classical key. If some properties of quantum physics can ensure that the choice cannot be detected and depended on Alice's inputs, then one-out-of-two QOT can be implemented on qubits directly, which means it will be simpler and more efficient.}
		\item \textbf{Others:} Besides the above, some QOT protocols have been proposed from different viewpoints or based on other problems that are difficlt for even a quantum computer to solve. Some researchers have proposed practical QOT\cite{Bennett1992c,Damgard2005,Li2014a}, etc., which are based on technological limitations. On the other hand, some researchers have turned the security assumptions into other problems too difficult for quantum computers. Yang et al.\cite{Yang2015b} proposed one-out-of-two QOT based on a symmetrically private information retrieval method in 2015; Souto et al.\cite{Souto2015} proposed all-or-nothing bit-string QOT based on $\mathrm{QSCD_{\mathit{ff}}}$\cite{Kawachi2012} and Rodrigues et al.\cite{Rodrigues2017} presented all-or-nothing bit-string QOT based on quantum public-key cryptography\cite{Nikolopoulos2008}.
	\end{enumerate}

\textbf{This research has 3 contributions. The first clearly arranges and decribes the complicated relationship among QOT, QBC and no-go theorems\cite{Mayers1997,Lo1997a,Lo1997}in the Introduction; the second discusses the relationship between the proposed protocol and Lo's no-go theorem\cite{Lo1997} in detail; while the third proposes a simple and efficient one-out-of-two QOT using nonorthogonal states, building up these states directly, rather than basing the protocol on all-or-nothing QOT, but otherwise corresponding to He's\cite{He2006} condition. The proposed protocol is therefore more efficient than the others discussed. The key point of the proposed protocol is that choice is made before Alice inputs her message $m_{0}$ and $m_{1}$. This makes one of two message to be the global phase, which cannot be measured and leaves obliviousness. This means that it is impossible for the proposed protocol to transfer more than one bit of information, achieving the purpose of one-out-of-two QOT.}
% under the Holevo's theorem\cite{Holevo1973}.

The remainder of this study is arranged as follows. Section 2 introduces the preliminaries, which include the quantum computation notation and some theorems. Section 3 illustrates the proposed protocol, and its relationship with no-go theorems\cite{Mayers1997,Lo1997a,Lo1997} in detail. Sections 4 and 5 analyze the proposed protocol's security and efficiency, and compare it to other protocols. Finally, conclusions are given in section 6.
The remainder of this study is arranged as follows. Section 2 introduces the preliminaries, which include the quantum computation notation and some theorems. Section 3 illustrates the proposed protocol, and its relationship with no-go theorems\cite{Mayers1997,Lo1997a,Lo1997} in detail. Sections 4 and 5 analyze the proposed protocol's security and efficiency, and compare it to other protocols. Finally, conclusions are given in section 6.
	
\section*{Results}\label{sec:results}
There are five subsections in this section, including the preliminaries, the proposed protocol, its relationship with Lo's no-go theorem\cite{Lo1997}, and its security and efficacy analysis. The preliminaries introduce the properties of quantum machines and key points of the proposed protocol. The protocol detail is decribed in the proposed protocol section. Its relationship with Lo's no-go theorem\cite{Lo1997} is discussed, with particular emphasis on the protocol's degree of compliance with He's proof\cite{He2006}. Finally, the security and efficacy analysis is decribed in the last two subsections.

	\subsection*{Preliminaries.}
	This subsection introduces the quantum machine notations and operations used in this study. The classical information carrier is called a ``bit''. The quantum information carrier is called a ``quantum bit'', or ``qubit''. A qubit collapses some states with a basis when it is measured. There are two common bases: the Z-basis and the X-basis. Those two bases are two bases of a 2D-plane where the Z-basis (standard basis) is $\ket{0}=\binom{1}{0}$ and $\ket{1}=\binom{0}{1}$ and the X-basis is $\ket{+}=\frac{1}{\sqrt{2}}\binom{1}{1}=\frac{1}{\sqrt{2}}(\ket{0}+\ket{1})$ and $\ket{-}=\frac{1}{\sqrt{2}}\binom{1}{-1}=\frac{1}{\sqrt{2}}(\ket{0}-\ket{1})$.%In usual, the bra and ket notation ($\bra{\cdot}$, $\ket{\cdot}$) are presented qubit state, and it is presents as row and column vector respectively. If $\left \langle \cdot ,\cdot \right \rangle=1$ and $\left \langle \blacksquare ,\cdot \right \rangle=0$,
	
	Two important quantum properties are superposition and entanglement. Superposition means that a qubit can simultaneously present $\ket{0}$ and $\ket{1}$ at the same time. For example, $\ket{\psi}=\alpha\ket{0}+\beta\ket{1}$, and this qubit will be collapsed to $\ket{0}$ and $\ket{1}$ with $\left |  \alpha\right |^{2}$ and $\left |  \beta\right |^{2}$ probabilities, respectively. For instance, state $\ket{-}$ is considered as superposition under the Z-basis. It has a $\left |  \frac{1}{\sqrt{2}}\right |^{2}=\frac{1}{2}$ probability of collapsing to $\ket{0}$, and a $\left |  -\frac{1}{\sqrt{2}}\right |^{2}=\frac{1}{2}$ probability of collapsing to $\ket{1}$.
	
	Moreover, unitary operations are considered as gates in quantum computers. There are four common operations, called Pauli matrices, which are $\{I, X, Y, Z\}$ as Eq.~\ref{eq:pauli}. As a result, operations $I$ and $Z$ are immunized at the Z-basis, and $I$ and $X$ are immunized at the X-basis. The results show that a single qubit cannot be observed for all four operations, which means some information is ignored, and this is a key element of the proposed protocol. For example, the state $\ket{0}$ after gate $Y$ becomes $-\ket{1}$, i.e. $\ket{1}$ will be given when the qubits is measured. This negative amplitude is called global phase, and cannot be measured. Another important gate is Hadamard gate as Eq.~\ref{eq:pauli}, also called $H$ gate. The $H$ gate can be used to convert between two different bases (Z- and X-basis). It corresponds to $HH^{*}=I$ and $H=H^{*}$. For example, the state $\ket{0}$ ($\ket{+}$) after gate $H$ becomes $\ket{+}$ ($\ket{0}$).
	
	Entanglement is another important property, and means that qubits cannot be presented as single. There are four common entangled states, called Bell states as Eq.~\ref{eq:bell}. For example, when measure $\ket{\Phi^{+}}$ is measured in Eq.~\ref{eq:bell}, the result may be either $\ket{00}_{12}$ or $\ket{11}_{12}$, where the subscript indicates the qubit order. As a result, it is possible to know two qubit states immediately when one is measured. Einstein referred to this as ``spooky action at a distance''.
	\begin{equation}\label{eq:pauli}
		I=\begin{pmatrix} 1 & 0 \\ 0 & 1 \end{pmatrix},\quad
		X=\begin{pmatrix} 0 & 1 \\ 1 & 0 \end{pmatrix},\quad
		Y=\begin{pmatrix} 0 & 1 \\ -1 & 0 \end{pmatrix},\quad
		Z=\begin{pmatrix} 1 & 0 \\ 0 & -1 \end{pmatrix},\quad
		H=\frac{1}{\sqrt{2}}\begin{pmatrix} 1 & 1 \\ 1 & -1 \end{pmatrix}
	\end{equation}
	\begin{equation}\label{eq:bell}
		\begin{split}
			\ket{\Phi^{+}}=\frac{1}{\sqrt{2}}(\ket{00}+\ket{11})_{12}, 
			\ket{\Phi^{-}}=\frac{1}{\sqrt{2}}(\ket{00}-\ket{11})_{12}, \\
			\ket{\Psi^{+}}=\frac{1}{\sqrt{2}}(\ket{01}+\ket{10})_{12}, 
			\ket{\Psi^{-}}=\frac{1}{\sqrt{2}}(\ket{01}-\ket{10})_{12}
		\end{split}
	\end{equation}
	
	\subsection*{The proposed protocol}\label{subsec:protocol}
	Consider the following scenario. First, is the one-side two-party protocol. Suppose Alice (sender) has a secret input $i\in\{1, 2, ..., n\}$ and Bob (receiver) has a secret input $j\in\{1, 2, ..., m\}$. Then Alice helps Bob to compute a function $f(i, j)\in\{1, 2, ..., p\}$. According to the above, the one-out-of-two OT can be mapped so that the Alice's input $i$ can be considered as her secret message $i(m_{0}, m_{1})$, and Bob's input $j$ can be considered as his choice. In addition, there are three security requirements for one-out-of-two OT: (a) Bob learns $f(i, j)$ unambiguously, (b) Alice learns nothing about $j$ and $f(i, j)$, and (c) Bob learns nothing about $i$ more than what logically follows from the values of $j$ and $f(i, j)$.
	
	The basis of the proposed protocol is B92\cite{Bennett1992}, which has been proved unconditionally secure both in theory and implementation\cite{Quan2002,Tamaki2004,Lucamarini2009}, which means that no one can identify all bases of the qubit, perfectly, without any information from the creator. Another key property is that some operations are immune at some bases, which means that it is not possible to identify all operations by single qubit. The proposed protocol can test Bob's loyalty under the security requirements; that is, Alice can check the initial states are correct. In addition, if Alice wants to lie to Bob, she will cause the error, which can be discovered by Bob at application level. The proposed protocol consists of six steps as follows:
	\begin{description}[leftmargin=3.5em]
		\item [Step 1.] Bob creates a candidate qubit sequence according to his choice intentions $j_{0}$ and $j_{1}$ to state $\ket{0}$ and $\ket{+}$, respectively. The $H$ gate can be considered as his choice intention in this stage. This sequence must be longer than the OT sequence containing the number of received messages, channel checking and Bob's loyalty testing qubits. In addition, the channel checking and loyalty testing states are different, the former belonging to $\{\ket{0}, \ket{1}, \ket{+}, \ket{-}\}$, where those qubits are also called decoy qubits, and the latter belonging to $\{\ket{0}, \ket{+}\}$. That is, $N$ is the minimum length at which Bob will receive $N$ messages, $M$ is the channel checking number, $K$ is the number of Bob's loyalty testing qubits, and the total length of the QOT sequence is $N+M+2K$. Bob reandomly prepares $M$ qubits from $\{\ket{0}, \ket{1}, \ket{+}, \ket{-}\}$ (each qubit is independent), and inserts them to the sequence. After this, he also inserts his $N$ and $2K$ candidate choice intentions ($\{\ket{0}, \ket{+}\}$) into the sequence, and then sends them to Alice.
		\item [Step 2.] Once she receives, the sequence from Bob, Alice first checks the channel for an eavedropper (Eve), and then tests Bob's loyalty. First, she asks Bob to publish the bases and states that he created. If the measurement results are greater that the given error rate, then Eve is present on the channel, and Alice and Bob abort their communication; otherwise, Alice then goes on to test Bob. She abandons qubits for channel checking and randomly selects some positions in order to request that Bob publish their bases. If other measurement results, i.e. $\notin\{\ket{0}, \ket{+}\}$, are greater than the given error rate, then Bob is considered dishonest, and she aborts this communication; otherwise, she proceeds to the next step.
		\item [Step 3.] Since the loyalty test may disturb the order of Bob's intentions, Bob must ask Alice to reorder those qubits. Twice $K$ qubits can fill up the vacancy of intentions. At this step, the sequence after reordering represents Bob's real choices.
		\item [Step 4.] Alice now inputs her secret message $m_{0}$ and $m_{1}$ by operations $I, Z, X, Y$ according to the combinations ``00'', ``01'', ``10'' and ``11'', respectively.
		\item [Step 5.] Alice then inserts the decoy qubits $\{\ket{0}, \ket{1}, \ket{+}, \ket{-}\}$ randomly in the sequence for channel checking, then sends them to Bob.
		\item [Step 6.] When Bob receives the sequence from Alice, he asks Alice to publish the positions and states of the decoy qubits. If the error rate is higher than channel error rate, they abort this communication and return to step 1. Otherwise, Bob learns the messages by the bases he prepared.
	\end{description}
	
	For a simple example of the proposed protocol, with only two received messages at the end of the protocol and without channel the checking version, Bob prepares two qubits in $\ket{0+}_{12}$ to represent his choices, and an additional two qubits in $\ket{0+}_{34}$ for loyalty testing. He then sends these four qubits to Alice. Once she receives a sequence in order $\ket{0+0+}_{1234}$, Alice asks Bob to publish the state of the fourth qubit, and measures it for comparison to the publishing and the measurement result for loyalty testing. Bob then asks Alice to reorder the remaining qubits following the order 21, and the states become $\ket{+0}_{21}$, with qubit 3 aborted. Alice performs $Z$ and $X$ according to her messages ``$01_{12}$'' and ``$10_{34}$'' at qubits 2 and 1, respectively. This makes the sequence $\ket{-1}_{21}$. She then sends the sequence back to Bob. Bob then performs X- and Z-basis measurements to learn the second and first messages, ``1'' and ``1'', respectively.
	
	\subsection*{Relationship with no-go theorems\cite{Mayers1997,Lo1997a,Lo1997}}\label{subsec:relationship}
	Two no-go theorems\cite{Mayers1997,Lo1997a,Lo1997} are discussed in this study: the MLC no-go theorem\cite{Mayers1997,Lo1997a} and Lo's no-go theorem\cite{Lo1997}. The MLC no-go theorem\cite{Mayers1997,Lo1997a} provides a strategy for cheating protocols similar to BB84\cite{Bennett1984}. This makes bit-commitment no-go. Lo's no-go theorem\cite{Lo1997} provides a strategy for learning all messages in one-side two-party secure computation, which means that one-out-of-two QOT is insecure.
	\begin{description}
		\item \textbf{MLC no-go theorem\cite{Mayers1997,Lo1997a}:} The MLC no-go theorem\cite{Mayers1997,Lo1997a} shows the entangled cheat, which means that the sender can create Eq.~\ref{eq:entangledcheat} and obtain the same results as the receiver. For example, the sender prepares the sequence under the qubit states $\{\ket{0}, \ket{1}, \ket{+}, \ket{-}\}$ and sends them to the receiver. The receiver then randomly selects the checking qubits, and asks the sender to publish the bases. As shown in Eq.~\ref{eq:entangledcheat}, the sender can determine the bases after the receiver's selections. The sender can then always escape the channel checking. This cheat strategy may steal all the messages involved. Indeed, the entangled strategy may be used as Dense Coding\cite{Bennett1992b}. If it is successful, Bob can learn all messages from Alice.
		
		However, the proposed protocol uses nonorthogonal states $\{\ket{0}, \ket{+}\}$, and thus resist to the entangled cheat strategy. Recall that in step 2 Alice asks Bob to publish the bases by her random choices, and because of Eq.~\ref{eq:entangledcheat}, the states $\ket{1}$ and $\ket{-}$ may be given, which means that each qubit for testing has a $\frac{1}{2}$ probability of detecting dishonest Bob. Therefore, the security $\xi$ can be determined by number of $K$ as Eq.~\ref{eq:entangledcheatsecurity}, and $N$ and $K$ are independent.
			\begin{equation}\label{eq:entangledcheat}
				\ket{\Phi^{+}}_{12}=\frac{1}{\sqrt{2}}(\ket{00}+\ket{11})_{12}=\frac{1}{\sqrt{2}}(\ket{++}+\ket{--})_{12}
			\end{equation}
			\begin{equation}\label{eq:entangledcheatsecurity}
				\xi=1-(\frac{1}{2})^{K}
			\end{equation}
		\item \textbf{Lo's no-go theorem\cite{Lo1997}:} The key point of Lo's no-go theorem\cite{Lo1997} is requirement (a), that ``Bob learns $f(i(m_{0}, m_{1}), j)$ unambiguously'', which leads to a 100\% that the selected state will collapse. In addition, the result is given after operations that are reversible. As a result, the choice can be made repeatedly to learn all messages. That is $f(i(m_{0}, m_{1}), j)\overset{U_{j_{0}}}{\rightarrow}f(i(m_{0}, m_{1}), j_{0})=m_{0}$, where $j$ represents not yet selected, Bob can perform inverse matrix $U_{j_{0}}^{-1}$ to obtain $m_{0}\overset{U_{j_{0}}^{-1}}{\rightarrow}f(i(m_{0}, m_{1}), j)$. Bob can learn all messages by repeating the above flow. However, He and Wang\cite{He2006} provided detailed discussion and definitions of two scenarios, namely ideal one-side two-party secure computation (definition A) and rigorous one-out-of-two OT (definition B). Obviously, definition B is a special case of definition A, and definition B is the scenarios of Lo's proof\cite{Lo1997}. As a result, definition B is that Alice inputs her messages first, then Bob inputs his choice. The important discussion of He and Wang\cite{He2006} is whether it is equivalent for Bob to input his choice first, and for Alice to subsequently input her message by Cr\'epeau's reduction\cite{Crepeau1988}? The answer is no: the funciton becomes $f(i(m_{0}, m_{1}, j), j)$ when Bob inputs his choice first, and $f(i(m_{0}, m_{1}, j_{0}), j_{1})$ is meaningless for $f(i(m_{0}, m_{1}, j), j)$. Therefore, he cannot change $i$ from $i(m_{0}, m_{1}, j_{0})$ to $i(m_{0}, m_{1}, j_{1})$. This proof shows that the order of choice and message input may be different, which means they are dependent. Extending the concept of He's proof\cite{He2006} to the general case\cite{He2015a}, if Alice and Bob should interact with each other and Bob cannot eliminate his operation independently, it does not belong to Lo's no-go theorem\cite{Lo1997}.
		
		In the proposed protocol (definition C), following the above relation, two conditions must be secure: (1) Bob's choice must correspond to He's proof\cite{He2006,He2015a}, and (2) the bases that Bob prepares cannot be known 100\%. To show (1), definitions C-i and C-ii, mean that the immune process and no one unitary operation can rotate two different qubit states (nonorthogonal) to be the same (if so, the operation becomes irreversible and cannot be used for a quantum machine). On the other hand, the global phase must also not be detectable. In other word, the choice can be considered as $H$, and message inputs can be considered as $\{I, X, Y, Z\}$. The result of $H\times G$ different from $G\times H$, where $G\in\{I, X, Y, Z\}$. As a result, Bob cannot eliminate his inputs without Alice's help, and it belongs to He's proof\cite{He2006,He2015a}. Therefore, these conditions show that the proposed function belongs to $f(i(m_{0}, m_{1}, j), j)$ and corresponds to He's proof\cite{He2006,He2015a}. (2) According to the B92\cite{Bennett1992}, Alice can obtain 25\% of Bob's choice, and the remaining 75\% remains unknown. For the remaining 75\% of qubits, Alice may prepare $\{\ket{0}, \ket{1}, \ket{+}, \ket{-}\}$ to replace the originals randomly. After this, Bob performs measurement according to the bases he prepared. That is, Alice cannot know which states Bob measured, as this would violate more than one conditions from definition A, and means that Alice can never cheat Bob because the error would be discovered at the application level. These discussions show that the proposed protocol is not a Lo's no-go theorem\cite{Lo1997}, and successfully builds one-out-of-two QOT.
		\begin{itemize}
			\item \textbf{Definition A: Ideal one-sided two-party secure computation}
				\begin{description}[leftmargin=3em]
					\item [(A-i)\hspace{1.5mm}] Bob learns $f(i, j)$ unambiguously.
					\item [(A-ii)\hspace{0.5mm}] Alice learns nothing about $j$ and $f(i, j)$.
					\item [(A-iii)] Bob learns nothing about $i$ more than what logically follows from the values of $j$ and $f(i, j)$.
				\end{description}
			\item \textbf{Definition B: rigorous one-out-of-two OT}
				\begin{description}[leftmargin=3em]
					\item [(B-i)\hspace{1.5mm}] Alice inputs $i$, which is a pair of message $(m_{0}, m_{1})$.
					\item [(B-ii)\hspace{0.5mm}] Bob inputs $j$ = 0 or 1.
					\item [(B-iii)] At the end of the protocol, Bob learns message $m_{j}$, but not the other message $m_{\bar{j}}$, i.e., the protocol is an ideal one-side two-party secure computation $f(m_{0}, m_{1}, j=0)=m_{0}$ and $f(m_{0}, m_{1}, j=1)=m_{1}$.
					\item [(B-iv)\hspace{0.5mm}] Alice does not know which $m_{j}$ Bob received.
				\end{description}
			\item \textbf{Definition C: The proposed protocol without testing}
				\begin{description}[leftmargin=3em]
					\item [(C-i)\hspace{1.5mm}] Bob inputs $j$ = 0 or 1 to change the qubit state to $\{\ket{0}, \ket{+}\}$ (Z- or X-basis) according to his choice intention.
					\item [(C-ii)\hspace{0.5mm}] Alice inputs her messages $m_{0}$ and $m_{1}$ using $\{I, X, Y, Z\}$.
					\item [(C-iii)] Bob learns either message $m_{0}$ and $m_{1}$ by the bases (Z- or X-basis) he prepared.
				\end{description}
		\end{itemize}
	\end{description}
	
	\subsection*{Security Analysis}\label{subsec:security_analysis}
	Two security conditions are considered in this study: external and internal attacks. External attacks involve an eavesdropper, Eve, attempting to steal messages without being detected. Internal attacks involve either Alice or Bob attempting to steal the other's secret, i.e., Alice wants to learn Bob's choices, or Bob wants to learn both of Alice's messages.
		\subsubsection*{External Attack}
		Alice and Bob must ensure that the communication channel between them is secure, because without channel checking or reduced frequency\cite{Fei2008}, Eve will be able to eavesdrop on their messages illicitly. In the proposed protocol, several single qubits are used as decoy qubits $\in\{\ket{0}, \ket{1}, \ket{+}, \ket{-}\}$, and randomly inserted into the transmitted sequence for channel checking, as in steps 1 and 5 of the protocol. The positions and states of those qubits are then published and measured in order to check whether Eve is present. If the measurement results with the same bases are different and the error rate is higher than the channel error rate, then Eve is present. Two common external attack strategies are the intercept-and-resend attack, and the entangling attack. They are discussed below:
		\begin{description}
			\item [\textbf{Intercept-and-resend attack:}] Eve intercepts all qubits and measures them in order to obtain the messages during transmission when the sender sends the qubit sequence to the receiver, and then resends those qubits to the receiver. This action should disturb the qubit states including decoy qubits. According the detection rate of BB84\cite{Bennett1984}, each qubit has $\frac{1}{4}$ probability of detecting Eve, and the detection rate increases with the number of decoy qubits $M$. As a result, the detection rate can be given by legal agents with the security $\xi$ as Eq.~\ref{eq:intercept_detection_rate}.
			\begin{equation}\label{eq:intercept_detection_rate}
				\xi=1-(\frac{3}{4})^{M}
			\end{equation}
			\item [\textbf{Entangling attack:}] Eve may use another method that does not disturb the qubit states, which is the entangling attack. She intercepts the transmission sequence, prepares an ancillary qubit $\ket{E}$ and performs a unitary operation $U_{e}$ on the intercepted qubit in order to entangle it with her qubit $\ket{E}$ during transmission. The unitary operation $U_{e}$ can be defined as follows:
			\begin{equation}\label{eq:entangled_attack}
				\begin{split}
					U_{e}(\ket{0}\ket{E})=a\ket{0}\ket{e_{00}}+b\ket{1}\ket{e_{01}}\\
					U_{e}(\ket{1}\ket{E})=c\ket{0}\ket{e_{10}}+d\ket{1}\ket{e_{11}}\\
					U_{e}(\ket{+}\ket{E})=\frac{1}{\sqrt{2}}(a\ket{0}\ket{e_{00}}+b\ket{1}\ket{e_{01}}+c\ket{0}\ket{e_{10}}+d\ket{1}\ket{e_{11}})\\
					=\frac{1}{2}\ket{+}(a\ket{e_{00}}+b\ket{e_{01}}+c\ket{e_{10}}+d\ket{e_{11}})+\frac{1}{2}\ket{-}(a\ket{e_{00}}-b\ket{e_{01}}+c\ket{e_{10}}-d\ket{e_{11}})\\
					U_{e}(\ket{-}\ket{E})=\frac{1}{\sqrt{2}}(a\ket{0}\ket{e_{00}}+b\ket{1}\ket{e_{01}}-c\ket{0}\ket{e_{10}}-d\ket{1}\ket{e_{11}})\\
					=\frac{1}{2}\ket{+}(a\ket{e_{00}}+b\ket{e_{01}}-c\ket{e_{10}}-d\ket{e_{11}})+\frac{1}{2}\ket{-}(a\ket{e_{00}}-b\ket{e_{01}}-c\ket{e_{10}}+d\ket{e_{11}}),
				\end{split}
			\end{equation}
			where $\ket{e_{00}}$, $\ket{e_{01}}$, $\ket{e_{10}}$, and $\ket{e_{11}}$ are four states determined by unitary operation $U_{e}$, and $\left | a \right |^{2}+\left | b \right |^{2}=1$ and $\left | c \right |^{2}+\left | d \right |^{2}=1$. If Eve wants to avoid detection, operation $U_{e}$ must satisfy $a=d=1$, $b=c=0$ and $\ket{e_{00}}=\ket{e_{11}}$, and as a result, no information can be obtained from the proposed protocol in this way.
		\end{description}
		
		\subsubsection*{Internal Attack}
		Internal attacks involve the legal agents Alice and Bob attempting to steal each other's secret information, i.e., Alice wants to learn Bob's choice, and Bob wants to learn all messages sent by Alice. Therefore, two conditions must be discussed, which are Alice's strategy and Bob's strategy:
		\begin{description}
			\item [\textbf{Alice's strategy:}] A dishonest Alice can obtain 25\% of Bob's choice, as with B92\cite{Bennett1992}, because the incorrect basis measurement can lead to incorrect measurement results. For example, Bob sends state $\ket{0}$ to Alice, the incorrect basis (X-basis) used means a $\frac{1}{2}$ probability, and incorrect state ($\ket{-}$) is obtained, which also has $\frac{1}{2}$ probability, and the total probability is $\frac{1}{4}=0.25$. The remaining 75\% is unknown, which means that Alice should randomly create some states $\in\{\ket{0}, \ket{1}, \ket{+}, \ket{-}\}$ to send to the receiver Bob. However, she cannot know Bob's final meausrement results because he does not publish any information about the bases. That is , he cannot decrypt correctly with $\frac{3}{4}\times\frac{1}{2}\times\frac{1}{2}=18.75\%$ for each bit at the application level, and will be aware that Alice is dishonest. For example, in a simple coin-flipping\cite{Bennett1984} protocol, Alice creates a random bit-string to present her coin-flipping result. She then communicates her bit-string to Bob with the proposed QOT. Bob publicly guesses the coin-flipping result. Alice then sends her random bit-string to Bob. He can compare the result with his QOT result. If Alice wants to learn Bob's choice, each bit has an 18.75\% of being wrong, Bob can thus detect that Alice is cheating.
			
			On the other hand, if the protocol\cite{Chou2010} works with a dummy message encrypted by the quantum resource, each qubit of the dummy message has an $18.75\%\times 50\%=9.375\%$ of being wrong, where the 50\% is the probability that the error state will result. 
			\item [\textbf{Bob's strategy:}] A dishonest Bob cdan prepare entangled qubit as an MLC no-go theorem\cite{Mayers1997,Lo1997a}, as in Eq.~\ref{eq:entangledcheatsecurity}. In this way, he can learn all messages from Alice, as in Eq.~\ref{eq:dense_coding}, where the subscript represents the qubit order, i.e., he can perfectly identify which operation Alice performed on qubit 1. However, only two states, $\ket{0}$ and $\ket{+}$, can be measured in the proposed protocol. Alice randomly selects $K$ positions, and asks Bob to publish the bases he prepared in step 2. If other measurement results are given, i.e. $\notin\{\ket{0}, \ket{+}\}$, Bob is dishonest. Furthermore, the detection rate shown in Eq.~\ref{eq:entangledcheatsecurity}, and the security $\xi$ can be determine by legal agents.
			\begin{equation}\label{eq:dense_coding}
				\begin{split}
					I_{1}\ket{\Phi^{+}}=I_{1}\frac{1}{\sqrt{2}}(\ket{00}+\ket{11})_{12}=\frac{1}{\sqrt{2}}(\ket{00}+\ket{11})_{12}=\frac{1}{\sqrt{2}}(\ket{++}+\ket{--})_{12}=\ket{\Phi^{+}}\\
					X_{1}\ket{\Phi^{+}}=X_{1}\frac{1}{\sqrt{2}}(\ket{00}+\ket{11})_{12}=\frac{1}{\sqrt{2}}(\ket{01}+\ket{10})_{12}=\frac{1}{\sqrt{2}}(\ket{++}-\ket{--})_{12}=\ket{\Psi^{+}}\\
					Y_{1}\ket{\Phi^{+}}=Y_{1}\frac{1}{\sqrt{2}}(\ket{00}+\ket{11})_{12}=\frac{1}{\sqrt{2}}(\ket{01}-\ket{10})_{12}=\frac{1}{\sqrt{2}}(\ket{+-}-\ket{-+})_{12}=\ket{\Psi^{-}}\\
					Z_{1}\ket{\Phi^{+}}=Z_{1}\frac{1}{\sqrt{2}}(\ket{00}+\ket{11})_{12}=\frac{1}{\sqrt{2}}(\ket{00}-\ket{11})_{12}=\frac{1}{\sqrt{2}}(\ket{+-}+\ket{-+})_{12}=\ket{\Phi^{-}}\\
				\end{split}
			\end{equation}
		\end{description}
	\subsection*{Efficacy Analysis}\label{subsec:efficacy_analysis}
\begin{figure}[ht]
		\centering
		\includegraphics[width=5in]{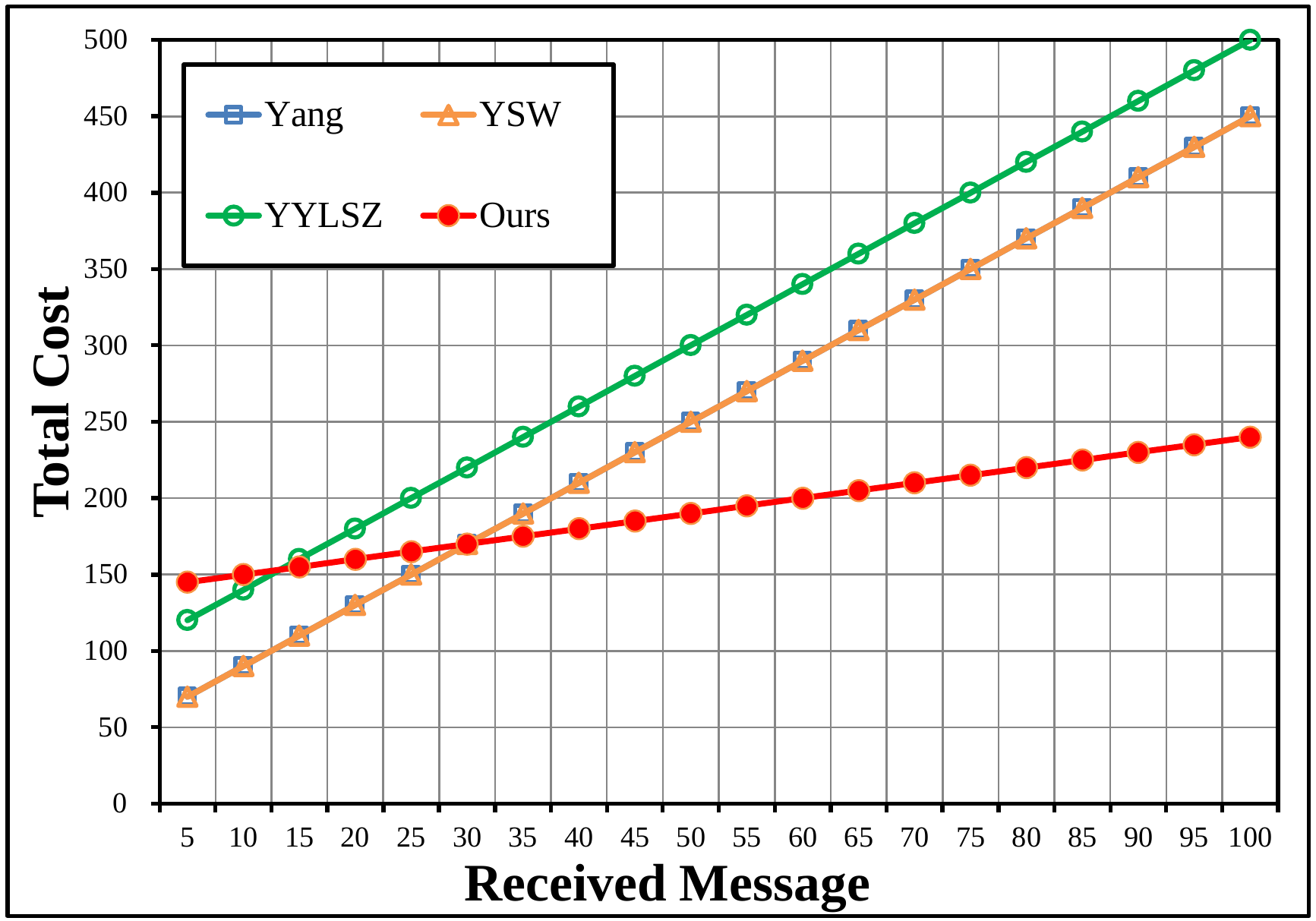}
		\caption{Illustration of the comparison results}
		\label{fig:comparison}
	\end{figure}

	In this section, the performance of the proposed protocol is compared with those of three modern one-out-of-two QOTs\cite{Yang2013,Yang2015b,Yang2015a} based on Cr\'epeau's reduction\cite{Crepeau1988}. These protocols are two-level structures that build one-out-of-two OT on all-or-nothing QOT. In addition, the probability of all-or-nothing QOT $p$ ($p$ is the probability of the unambiguous key) is not always 50\%. It requires significant quantum resources to build two classical keys (one is unambiguous and the other is unknown) using all-or-nothing QOT for one-out-of-two OT. In addition, every transmission should include decoy qubits for channel checking. Some protocols may need many transmissions to complete all-or-nothing QOT, and use many decoy qubits. For fairness, the variable security $\xi$ is at least 99.9999\% as long as 50 decoy qubits are used for each transmission. After this, the most important indicators are the conversion efficiency of two OTs, number of transmissions (which indirectly affects the number of decoy qubits) and loyalty testing. The total cost of all protocols under the $R$ messages received requirement is calculated, and results are given in Table ~\ref{tab:comparison}. Here, only the quantum cost is calculated, which means that only the all-or-nothing QOT part of other protocols is counted. Detailed descriptions of these protocols are given below:
	\begin{description}
		\item [\textbf{Yang's protocol\cite{Yang2013}:}] This protocol uses the B92\cite{Bennett1992} protocol as the all-or-nothing QOT base. Therefore, it requires four qubits for an unambiguous key and only one transmission. However, this research focuses on the bit-commitment protocol more than the OT protocol, with no further security analysis of the QOT protocol, or strategies for detecting Eve. Therefore, no strategy is provided for the loyalty testing between Alice and Bob. In addition, the number of decoy qubits is computed as defined above because the orginal detecting strategy of B92\cite{Bennett1992} is mroe inefficient. As a result, the total quantum cost is $4\times R+50$.
		\item [\textbf{YSW protocol\cite{Yang2015b}:}] This protoocl is the reduction of BB84\cite{Bennett1984} to B92\cite{Bennett1992}. It uses the BB84\cite{Bennett1984} strategy and publishes additional state information in order to make Bob generate unambiguous keys, as in B92\cite{Bennett1992}. It also requires four qubits for an unambiguous key. In addition, it takes 1 transmission to complete the all-or-nothing QOT. However, it does not include a loyalty testing method for the all-or-nothing QOT element. Errors may be detect at application level. The overall cost of the protocol is $4\times R+50$.
		\item [\textbf{YYLSZ protocol\cite{Yang2015a}:}] This protocol also requires at least four qubits to obtain an unambiguous key using its all-or-nothing QOT strategy, i.e. $p=\frac{1}{4}$. In the all-or-nothing QOT protocol, Alice first sends a sequence to Bob, and Bob then sends it back after hsi measurement, which takes two transmissions. After this, Alice can test Bob's loyalty by observing the probability of occurrence of states $\ket{+}$ and $\ket{-}$. Note that the strategy must be based on a large number of qubit consumptions. On the other hand, Bob cannot really test Alice's loyalty, and is only able to detect errors in the later application by one-out-of-two OT. The overall cost of the protocol is $4\times R+2\times50$.
		\item [\textbf{Proposed protocol:}] The proposed protocol requires one qubit for one bit receiving. Alice requires $K=20$ qubits (each qubit for testing has a $\frac{1}{2}$ probability of detecting dishonest Bob) to test Bob's loyalty with a 0.\\99.9999\% accuracy. As a result, Bob should prepare a total of $2K=40$ qubits (half $K$ is the alternatives of intentions) to ensure the vacancy of intentions. In addition, the proposed protocol is able to ensure the loyalty of both agents because Alice can test it by the nonorthogonal state properties, and Bob can detect the error at application level. The overall cost of the proposed protocol is $R+2\times50+2\times20$.
	\end{description}
	From Table~\ref{tab:comparison}, the proposed protocol becomes most efficient when $R\geqslant30$, as shown in Figure~\ref{fig:comparison}. This demonstrates that building one-out-of-two QOT directly is more efficient than Cr\'epeau's reduction\cite{Crepeau1988}.
	\begin{table}[ht]
		\centering
		\begin{tabular}{|l|l|l|l|l|l|l|}
			\hline
			1-2 QOT Protocols & \tabincell{l}{\textsuperscript{1}Quantum Resource\\for a Message} & \tabincell{l}{\textsuperscript{2}Transmission\\Times} & \tabincell{l}{\textsuperscript{3}Decoy\\Qubits} & \tabincell{l}{\textsuperscript{4}Loyalty\\Testing Qubits} & \tabincell{l}{\textsuperscript{5}Loyalty\\Testing} & \tabincell{l}{\textsuperscript{6}Total Cost} \\
			\hline
			Yang's protocol\cite{Yang2013} & $4$ & 1 & $50$ & 0 & None & $4\times R+50$ \\
			\hline
			YSW protocol\cite{Yang2015b} & 4 & 1 & $2\times50$ & 0 & None & $4\times R+50$ \\
			\hline
			YYLSZ protocol\cite{Yang2015a} & 4 & 2 & $2\times50$ & 0 & Alice & $4\times R+2\times50$ \\
			\hline
			Our protocol & 1 & 2 & $2\times50$ & $2\times20$ & Alice & $R+2(50+20)$ \\
			\hline
			\multicolumn{7}{l}{\textsuperscript{1}\footnotesize{Quantum Resource for a message: Number of quantum resources consumed for receiving a bit without decoy qubits.}}\\
			\multicolumn{7}{l}{\footnotesize{\textsuperscript{2}Transmission Times: Number of transmissions for a sequence.}}\\
			\multicolumn{7}{l}{\footnotesize{\textsuperscript{3}Decoy Qubits: Number of decoy qubits, considering the transmission times.}}\\
			\multicolumn{7}{l}{\footnotesize{\textsuperscript{4}Loyalty Testing Qubits: Number of loyalty testing qubits.}}\\
			\multicolumn{7}{l}{\footnotesize{\textsuperscript{5}Loyalty Testing: Which parties are able to test the loyalty of the other.}}\\
			\multicolumn{7}{l}{\footnotesize{\textsuperscript{6}Total Cost: The total quantum consumption under $R$ bits received requirement.}}
		\end{tabular}
		\caption{\label{tab:comparison}Comparison of the performance of three modern one-out-of-two QOT with that of the proposed protocol.}
	\end{table}
\section*{Discussion}\label{sec:discussion}
This research successfully proposed a simple and efficient one-out-of-two QOT with single nonorthogonal qubits based on the most basic properties of quantum machines, namely the immunity process, the global phase and nonorthogonal states. Many important quantum machine applications\cite{Deutsch1992,Bennett1992,Grover1996} are based on these properties. If one of these properties fails, many applications will become invalid, the basis of quantum machines will be unstable. In addition, reordering is applied to probide security against internal attacks by dishonest Alice, while dishonest Bob can always be detected at the application level because dishonest Alice may cause the errors. As a result, the proposed protocol is not onlyh easily implemented, but is also more efficient than the compared protocols\cite{Yang2013,Yang2014,Yang2015a,Yang2015c,Yang2017} based on classical Cr\'epeau's reduction\cite{Crepeau1988}.

%\section*{Acknowledgements (not compulsory)}
%\label{sec:acknowledgments}
%Acknowledgements should be brief, and should not include thanks to anonymous referees and editors, or effusive comments. Grant or contribution numbers may be acknowledged.

\bibliography{main}

\section*{Author contributions statement}\label{sec:authors}
Y.-H. Chou designed the scheme. G.-J. Zeng and Z.-H. Chang developed method. Y.-S. Yang did security analysis and consumption comparison. All authors edited and reviewed the manuscript.

\section*{Additional information}\label{sec:additional}
\textbf{Competing Interests:} The authors declare no competing financial interests.

%The corresponding author is responsible for submitting a \href{http://www.nature.com/srep/policies/index.html#competing}{competing financial interests statement} on behalf of all authors of the paper. This statement must be included in the submitted article file.

%figure should be placed here

%table should be placed here

%Figures and tables can be referenced in LaTeX using the ref command, e.g. Figure \ref{fig:stream} and Table \ref{tab:example}.

\end{document}